\documentclass[twocolumn,floatfix,showpacs,prb,10pt,superscriptaddress]{revtex4-1}
\usepackage{graphicx}
\usepackage{amsmath}
\usepackage{amssymb}
\usepackage{bm}
\usepackage{hyperref}
\usepackage{multirow}
\usepackage{color}

\begin{document}

\title{Extended topological group structure due to average reflection symmetry}
\author{M. Diez}
\affiliation{Instituut-Lorentz, Universiteit Leiden, P.O. Box 9506, 2300 RA Leiden, The Netherlands}

\author{D. I. Pikulin}
\affiliation{Instituut-Lorentz, Universiteit Leiden, P.O. Box 9506, 2300 RA Leiden, The Netherlands}
\affiliation{Department of Physics and Astronomy, University of British Columbia, Vancouver, BC, Canada V6T 1Z1 and Quantum Matter Institute, University of British Columbia, Vancouver BC, Canada V6T 1Z4}

\author{I. C. Fulga}
\affiliation{Instituut-Lorentz, Universiteit Leiden, P.O. Box 9506, 2300 RA Leiden, The Netherlands}
\affiliation{Department of Condensed Matter Physics, Weizmann Institute of Science, Rehovot 76100, Israel}

\author{J. Tworzyd{\l}o}
\affiliation{Institute of Theoretical Physics, Faculty of Physics, University of Warsaw,
Pasteura 5, 02--093
 Warsaw, Poland}

\date{\today}
\begin{abstract}
We extend the single-particle topological classification of insulators and superconductors to include systems in which disorder preserves average reflection symmetry. We show that the topological group structure of bulk Hamiltonians and topological defects is exponentially extended when this additional condition is met, and examine some of its physical consequences. Those include localization-delocalization transitions between topological phases with the same boundary conductance, as well as gapless topological defects stabilized by average reflection symmetry.
\end{abstract}
% \pacs{...}
\maketitle

\section{Introduction}
\label{sec:intro}

% TI have gapped bulk and gapless boundary -> invariants -> TI table
Topological insulators (TI) are states of matter in which the bulk is gapped, but which host protected gapless edge states.\cite{Hasan2010, Qi2011} This behavior was first studied in connection to the quantum Hall effect,\cite{Halperin1982,Buettiker1988} a two-dimensional system, and later generalized to include arbitrary dimensions, as well as boundary states protected by the fundamental symmetries of the system: time-reversal ${\cal T}$, particle-hole ${\cal P}$, and chiral symmetry ${\cal C}$.\cite{Kitaev2009, Schnyder2009} In each case, the gapless nature of boundary states is a consequence of the system's bulk properties. This enables obtaining topological invariants, quantities determined from the bulk which count the number of protected states at a termination of the system.
For single-particle systems, the group structure of topological invariants ($\mathbb{Z}$ or $\mathbb{Z}_2$) is listed in the so-called periodic table of topological insulators, which shows that in any dimension 5 out of the 10 Altland-Zirnbauer\cite{Altland1997} (AZ) symmetry classes can be topologically non-trivial. As long as the protecting symmetries are not broken, the invariant cannot change without closing the bulk gap, explaining the robustness of the boundary states to perturbations such as disorder.

% TI + lattice symmetry = WTI
Topologically non-trivial behavior can occur also due to symmetries of the underlying lattice. This enables weak and crystalline topological insulators in the presence of translational symmetry, or point group symmetries (rotation, reflection, etc.).\cite{Moore2007, Roy2009, Fu2007a, Fu2011, Hsieh2012} Many generalizations of the periodic table have been considered by examining the interplay between ${\cal T}$, ${\cal P}$, ${\cal C}$, and different lattice symmetries.\cite{Slager2012, Jadaun2013, Chiu2013, Alexandradinata2014, Teo2013, Zhang2013, Benalcazar2014, Morimoto2013, Chiu2014}

\begin{table}[!t]
\caption{Group structure of single-particle topological invariants in the ten AZ symmetry classes, with average reflection symmetry preserved along all directions. The strong invariants of the original TI table are shown in blue and those protected by ARS in black.\label{tab:ti}}
\begin{ruledtabular}
\begin{tabular}{ c | c c c }
 \multirow{2}{*}{Symmetry class} & & {Dimension} & \\
  & 1 & 2 & 3 \\
 \hline
 A    &  & \textcolor{blue}{$\mathbb{Z}$} & ${\mathbb{Z}_2}^3$ \\
 AIII &  \textcolor{blue}{$\mathbb{Z}$} & ${\mathbb{Z}_2}^2$ & \textcolor{blue}{$\mathbb{Z}$}$\times{\mathbb{Z}_2}^3$ \\
 \hline
 AI   &  &  &  \\
 BDI  & \textcolor{blue}{$\mathbb{Z}$} & ${\mathbb{Z}_2}^2$ & ${\mathbb{Z}_2}^3$ \\
 D    & \textcolor{blue}{$\mathbb{Z}_2$} & \textcolor{blue}{$\mathbb{Z}$}$\times{\mathbb{Z}_2}^2$ & ${\mathbb{Z}_2}^6$ \\
 DIII & \textcolor{blue}{$\mathbb{Z}_2$} & \textcolor{blue}{$\mathbb{Z}_2$}$\times{\mathbb{Z}_2}^2$ & \textcolor{blue}{$\mathbb{Z}$}$\times{\mathbb{Z}_2}^6$ \\
 AII  &  & \textcolor{blue}{$\mathbb{Z}_2$} & \textcolor{blue}{$\mathbb{Z}_2$}$\times{\mathbb{Z}_2}^3$ \\
 CII  & \textcolor{blue}{$\mathbb{Z}$} & ${\mathbb{Z}_2}^2$ & \textcolor{blue}{$\mathbb{Z}_2$}$\times{\mathbb{Z}_2}^3$ \\
 C    &  & \textcolor{blue}{$\mathbb{Z}$} & ${\mathbb{Z}_2}^3$ \\
 CI   &  &  & \textcolor{blue}{$\mathbb{Z}$} \\
\end{tabular}
\end{ruledtabular}
\end{table}

% WTI + disorder = STI
Disorder breaks all symmetries of the lattice, leading to a distinction between \emph{strong} and \emph{weak} topological insulators (WTI) and their associated invariants. Despite owing their protection to lattice symmetries, the boundary states of some WTIs may still survive disorder. This was first shown for a stack of quantum spin-Hall layers,\cite{Ringel2012, Mong2012, Fu2012} a three-dimensional WTI belonging to symmetry class AII in the AZ classification, and later generalized to systems of different dimensionality and symmetry class, dubbed statistical topological insulators.\cite{Fulga2014} Here, protection is not given by an exact symmetry, but by one which only holds on average. Whereas the original invariants belong to $\mathbb{Z}$ or $\mathbb{Z}_2$, those stabilized by average symmetries only have a $\mathbb{Z}_2$ group structure.

Motivated by the robustness of boundary states in statistical topological insulators, we study how the classification of TIs and topological defects are extended by average symmetries. For concreteness, we will focus on disordered systems which preserve average reflection symmetry (ARS), a situation which occurs in many condensed matter systems.\cite{Tanaka2012, Dziawa2012, Xu2012} Each element of the disordered ensemble of Hamiltonians, $H$, appears with equal probability as its reflected counter part, ${\cal R}_j^{-1} H {\cal R}_j$, with ${\cal R}_j$ a unitary reflection operator about the $j$-direction. Oblique reflection gives the same physics as the ordinary one, thus in the examples we will consider only the ordinary one. For us the relevant cases are when the reflection plane passes through a lattice site of the system, such that the symmetry can be broken by staggering the strength of consecutive hopping amplitudes.

We find that the group structure of topological invariants is exponentially enlarged by ARS, since weak invariants of all dimensions $d>0$ contribute simultaneously and independently to the classification presented in Table~\ref{tab:ti}. Some of the physical consequences of this extension include the possibility of disordered topological phase transitions governed only by a change in the weak invariant. We find a particularly interesting situation when the system possesses a nonzero strong index on both sides of such a transition.  Then the conductance of the boundary is non-trivial and identical in both phases, while at the transition the  bulk gap must close in the presence of ARS. Additionally, we show that the extended classification applies also to topological defects.\cite{Teo2010} It allows us to define a new class of gapless \emph{statistical topological defects}, which are robust to disorder but can only exist in the presence of average symmetries.

In the following, we begin our discussion by motivating the need for an extended topological classification with some concrete examples. In Section \ref{sec:model_D} we introduce a model for a two-dimensional (2d) topological superconductor in symmetry class D, exhibiting disordered phase transitions across which the strong invariant remains constant, and only a weak index changes. To show how this behavior escalates in higher dimensions, we consider a three-dimensional topological superconductor (class DIII) in Section \ref{sec:model_DIII}. Its disordered phases are distinguished by a second generation weak index, \emph{i.e.}~one which is two dimensions lower than the system dimension, even if the strong and 2d weak invariants don't change. We generalize these results to arbitrary dimension and symmetry class in Section \ref{sec:stpt} showing that ARS enlarges the topological classification of both bulk Hamiltonians and topological defects alike. We conclude in Section \ref{sec:conc}.

\section{Topological superconductor in class D}
\label{sec:model_D}

\begin{figure*}[htb]
 \includegraphics[width=2.0\columnwidth]{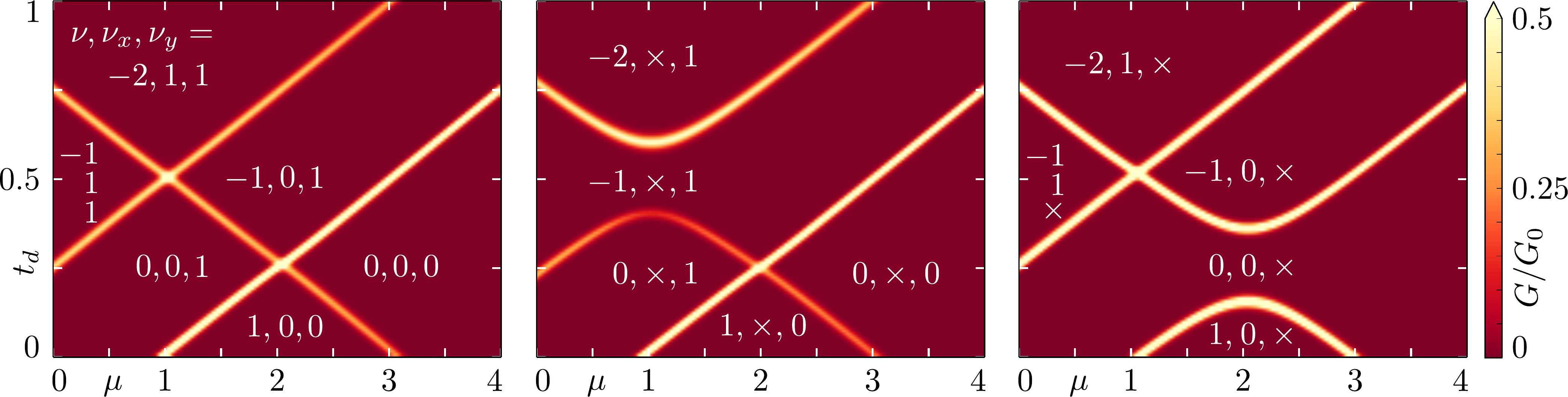}
 \caption{Bulk thermal conductance of a disordered system with Hamiltonian \eqref{eq:HclassD} as a function of $\mu$ and $t_d$. Parameters are $t_x=1$, $t_y=1/2$, $\Delta_x=2$, $\Delta_y=1$, and disorder strength $U=1$. Each phase is labeled according to its strong and weak topological invariants: $\nu,\nu_x,\nu_y$. With average reflection symmetry (left panel) the crossings are protected by the weak invariants. Breaking ARS in either the $x$- or the $y$-directions destroys the corresponding invariant (marked with $\times$) and leads to an anticrossing, as shown in the middle and right panels. In the middle panel the staggering strength in the $x$-direction is $s_x=0.2$, while in the right panel the $y$-direction hoppings are staggered with $s_y=0.4$. \label{fig:pd_D}}
\end{figure*}

Two-dimensional superconductors with broken time-reversal as well as spin-rotation symmetry belong to symmetry class D in the AZ classification. The minimal topological model is a $2\times 2$ Bogoliubov-De Gennes Hamiltonian describing spinless fermions in the presence of a $p$-wave order parameter, $\Delta({\bf k}) \sim {\bf k}$. The only constraint is provided by the particle-hole symmetry, and reads:

\begin{equation}
 \tau_x H({\bf k}) \tau_x = - H^*(-{\bf k}), \label{eq:phs}
\end{equation}
in terms of the Pauli matrices $\tau_i$ acting on the particle-hole degree of freedom.

We use a tight binding Hamiltonian of the form
\begin{equation}\label{eq:HclassD}
 H({\bf k})  = \varepsilon({\bf k}) \tau_z + \Delta_x \tau_x \sin(k_x) + \Delta_y \tau_y \sin(k_y),
\end{equation}
with
\begin{eqnarray}\label{eq:epsilonk}
 \varepsilon({\bf k}) &=& -2 t_x \cos(k_x) - 2 t_y \cos(k_y) - \mu \nonumber\\
 & & - 2 t_d \cos(k_x+k_y) - 2 t_d \cos(k_x-k_y).
\end{eqnarray}

Here, $\Delta_{x,y}$ is the strength of the $p$-wave pair potential, $t_{x,y}$ are the anisotropic hopping amplitudes in the $x$- and $y$-directions, and $\mu$ is the chemical potential. The Hamiltonian \eqref{eq:HclassD} is discretized on a square lattice of $L_x\times L_y=50\times 50$ sites (lattice constant $a=1$), with the last two terms of Eq.~\eqref{eq:epsilonk} leading to next nearest neighbor hoppings, parametrized by the diagonal hopping amplitude $t_d$. Disorder is modeled by random variations of the chemical potential, drawn independently for each site from the uniform distribution $\left[ \mu-U, \mu+U \right]$. In the following we set $t_x=1$ and express all other Hamiltonian parameters relative to this energy scale. All tight binding simulations are performed using the Kwant code.\cite{Groth2014}

We attach disorder free leads at $x=0, L_x$ connecting the system to reservoirs at temperatures $T_0$ and $T_0 + \delta T$. The Fermi level ($E=0$) scattering matrix,

\begin{equation}\label{eq:smatrix}
 S=\begin{pmatrix}
    r & t \\
    t' & r'
   \end{pmatrix},
\end{equation}
enables us to compute the thermal conductance $G=G_0\,{\rm Tr}\,t^{\dagger}t$, $G_0=\pi^2k_{\rm B}^2T_0/6h$, in the low-temperature, linear response regime, as well as the topological invariants of the system. The Chern number, the strong topological invariant of the system, reads\cite{Fulga2012, Fulga2011}

\begin{equation}\label{eq:d_invariant}
 \nu = \frac{1}{2\pi i}\int_0^{2\pi} d\phi \,\frac{d}{d\phi}\,\ln\,{\rm det}\,r(\phi),
\end{equation}
while the weak $\mathbb{Z}_2$ invariants are given by

\begin{equation}\label{eq:d_invariant_weak}
 (-1)^{\nu_y} = {\rm sign}\,{\rm det}\,r(\phi=0).
\end{equation}

In Eqs.~\eqref{eq:d_invariant} and \eqref{eq:d_invariant_weak} $r(\phi)$ is the reflection block of the scattering matrix in the presence of twisted boundary conditions applied to the states in the $y$ direction: $\psi(x,0)=e^{i\phi}\psi(x,L_y)$. The weak invariant in the $x$-direction is evaluated in a similar fashion, by attaching leads in the $y$-direction and using periodic boundary conditions ($\phi=0$) along $x$. Both the strong and the weak invariant is defined such that $\nu,\,\nu_y=0$ is trivial, while phases with non-zero invariants are non-trivial, either in the strong or weak sense.

As a function of $\mu$ and $t_d$, the system shows a variety of topological phases separated by phase transitions at which the bulk gap closes (see Fig.~\ref{fig:pd_D}, left panel). The phases are strong topological insulators whenever the Chern number is nonzero, with chiral Majorana zero modes on all edges. When $\nu=0$, we also find weak topological insulators, where two out of four edges avoid localization in the presence of disorder, hosting counter-propagating Majorana edge modes -- so-called Kitaev edges.\cite{Diez2014}

While typically the Chern number changes across a phase transition, in the model \eqref{eq:HclassD} there are also transitions across which the strong invariant remains constant, and only the weak invariants change. They are the crossings in Fig.~\ref{fig:pd_D}, occurring at $(\mu, t_d) = (1, 1/2)$ and $(2, 1/4)$. At $t_d=1/4$, varying the chemical potential causes a change of the weak invariant $\nu_y$, while the other weak invariant, $\nu_x$, is responsible for the phase transition at $t_d=1/2$. The bulk gap is closed at $(\mu, t_d)=(2, 1/4)$ even though there are the same number of chiral Majorana edge modes with the same chirality both for $\mu<2$ and $\mu>2$.

In the clean case ($U=0$) these anomalous topological phase transitions are protected by the exact reflection symmetry of the system.
We find in our simulations
that they persist when disorder is added, up to values of $U$ comparable to the bulk gap, when a thermal metal phase develops.\cite{Senthil2000, Evers2008, Medvedyeva2010} Note that in Fig.~\ref{fig:pd_D} we plot the bulk thermal conductance of a \emph{single} system at strong disorder, showing that at large enough system sizes ARS can
protect not only the
properties of the disordered ensemble as a whole, but its individual elements as well. The presence of crossings in the disordered phase diagram of Hamiltonian \eqref{eq:HclassD} shows that the Chern number, a $\mathbb{Z}$ index, is insufficient to describe class D two-dimensional disordered superconductors with ARS. The full topological classification is in fact $\mathbb{Z}\times{\mathbb{Z}_2}^2$.

We verify this group structure by selectively removing average symmetries from the system. This is done by staggering the $x$- and/or $y$-direction hoppings as $t_{x,y}\rightarrow t_{x,y} (1 + (-1)^{x, y} s_{x,y})$. For $s\neq 0$, consecutive hoppings in the same direction have alternating strength, such that ARS no longer holds. Breaking either of the average symmetries removes the protection of the associated weak invariant, and therefore splits the corresponding crossing, as shown in the middle and right panels of Fig.~\ref{fig:pd_D}. This signals that the two average symmetries act independently, justifying the extended $\mathbb{Z} \times {\mathbb{Z}_2}^2$ group structure.

\section{Topological superconductor in class DIII}
\label{sec:model_DIII}

\begin{figure*}[tb]
  \begin{center}
	 \includegraphics[width=2.0\columnwidth]{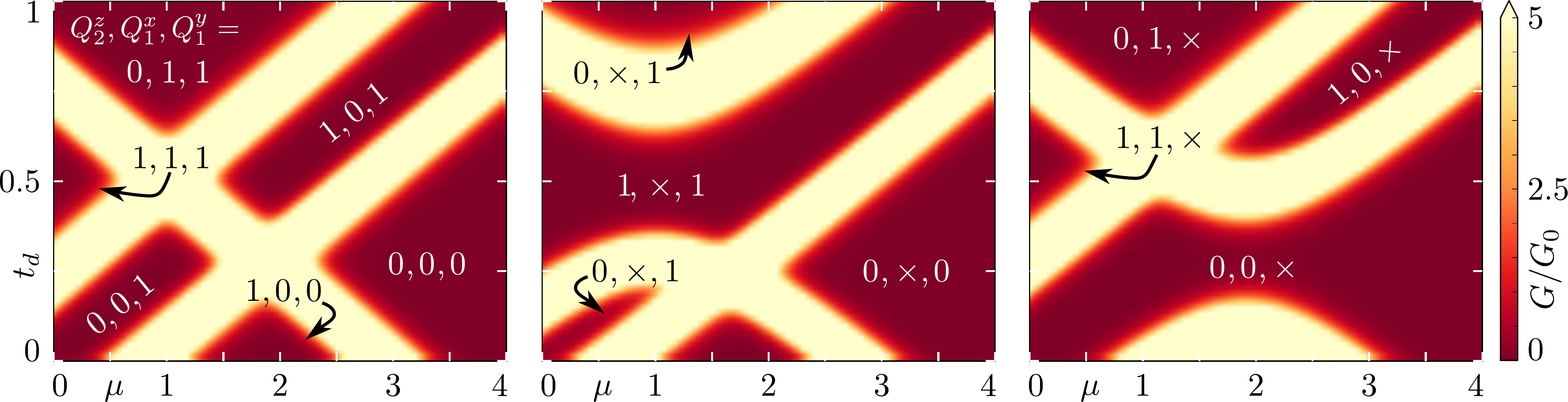}
  \end{center}
  \caption{Bulk thermal conductance of a single disordered system with Hamiltonian \eqref{eq:HDIII} as a function of $\mu$ and $t_d$. Parameters are $t_x=1$, $t_y=1/2$, $t_z=0.05$, $\Delta_x=3$, $\Delta_y=1.5$, $\Delta_z=0.15$, $K=0.2$, and disorder strength $U=1$. Phases are labeled by their topological invariants $Q_2^z$, $Q_1^x$, and $Q_1^y$, with $\times$ marking an invariant destroyed by breaking ARS. In the absence of staggering, phases are distinguished by both first and second generation weak invariants (left panel). Staggering in the $x$- and $y$-directions are set to $s_x=0.25$ in the middle panel and $s_y=0.5$ in the right panel, respectively.}
  \label{fig:DIII_crossing}
\end{figure*}

To demonstrate the protection of an insulating phase by a second generation weak invariant, \emph{i.e.}~an invariant two dimensions lower than the system dimension, we choose a model in symmetry class DIII, with Hamiltonian
\begin{align}
  H(\bm k) &= \varepsilon(\bm k)\sigma_0\otimes\tau_z + K\sigma_y\otimes\tau_y + \Delta_x\sin(k_x)\sigma_z\otimes\tau_x \nonumber\\
  & \quad + \Delta_y\sin(k_y)\sigma_o\otimes\tau_y + \Delta_z\sin(k_z)\sigma_x\otimes\tau_x,
  \label{eq:HDIII}
\end{align}
where
\begin{align}
  \varepsilon(\bm k) &= -2t_x\cos(k_x)-2t_y\cos(k_y)-2t_z\cos(k_z)\nonumber\\
	& \quad -2t_d\cos(k_x+k_y)-2t_d\cos(k_x-k_y).
\end{align}
The Pauli matrices $\tau_i$ and $\sigma_i$ act on the particle-hole and spin degree of freedom, respectively. Here, $t_{x,y,z}$ and $\Delta_{x,y,z}$ are the anisotropic hopping amplitudes and the $p$-wave pairing amplitudes in the $x$-, $y$-, and $z$-directions (as before, we set $t_x=1$). The chemical potential is $\mu$, while $K$ models an $s$-wave order parameter coupling the two spin blocks.
The model is constrained by particle-hole and time-reversal symmetry:
\begin{align}
  \tau_x H(\bm k)\tau_x &= -H^*(-\bm k), \\
  \sigma_y H(\bm k)\sigma_y &= H^*(-\bm k).
\end{align}
Like in the previous model, we introduce disorder by random spatial variations of the chemical potential, with disorder strength $U$.
We discretize the Hamiltonian \eqref{eq:HDIII} on a cubic lattice of linear size $L_{x,y,z}= 16$. Ideal leads are attached along one direction, and twisted boundary conditions are imposed in the other two, as $\psi(0, y, z) = e^{i\phi_x}\psi(L_x, y, z)$, $\psi(x, 0, z) = e^{i\phi_y}\psi(x, L_y, z)$, or $\psi(x, y, 0) = e^{i\phi_z}\psi(x, y, L_z)$. In each case the reflection matrix is a function of two out of the three twist angles $\phi_{x,y,z}$. Owing to time-reversal symmetry, the reflection block can be brought to an anti-symmetric form whenever the twist angles are $0$ or $\pi$ (periodic or anti-periodic boundary conditions), making its Pfaffian, ${\rm Pf}\, r$, well defined.
As in the class D model, the system shows different disordered topological phases as a function of $\mu$ and $t_d$, protected by 1d or 2d weak invariants (see Fig.~\ref{fig:DIII_crossing}). The relevant two-dimensional weak index reads\cite{Fulga2012, Fulga2011}
\begin{align}\label{eq:q_diii_2z}
\begin{split}
 (-1)^{Q^z_2} = {\rm sign}& \left[ {\rm Pf}\, r(\phi_y=0, \phi_z=0) \right. \times \\
 & \left.\, {\rm Pf}\, r(\phi_y=\pi, \phi_z=0) \right],
\end{split}
\end{align}
and is responsible for gapless modes on all side surfaces, \emph{i.e.}~surfaces parallel to the $z$-direction. Non-trivial 1d weak invariants appearing in Fig.~\ref{fig:DIII_crossing} are
\begin{equation}\label{eq:q_diii_1x}
 (-1)^{Q^x_1} = {\rm sign} \left[ {\rm Pf}\, ir(\phi_x=0, \phi_z=0) \right],
\end{equation}
and
\begin{equation}\label{eq:q_diii_1y}
 (-1)^{Q^y_1} = {\rm sign} \left[ {\rm Pf}\, ir(\phi_y=0, \phi_z=0) \right],
\end{equation}
leading to protected gapless modes on side surfaces parallel to the $x$- and $y$-directions, respectively. Three-dimensional class DIII systems also allow for a strong invariant, but this one remains zero throughout the phase diagram of Fig.~\ref{fig:DIII_crossing}, since the top and bottom surfaces are insulating whenever the bulk is gapped.

Unlike the two-dimensional model of Section \ref{sec:model_D}, in which topologically different phases were separated by insulator-to-insulator phase transitions, the three dimensional Hamiltonian \eqref{eq:HDIII} has finite-extent metallic regions.\cite{Fulga2012a} Nevertheless, insulating phases are not connected in the presence of ARS. We find that the weak 1d and 2d invariants are robust, leading to surfaces which do not localize once disorder is added. Breaking average reflection symmetry by staggering consecutive hoppings in the $x$- or $y$-directions destroys the corresponding invariants, connecting the phases as shown in the middle and right panels of Fig.~\ref{fig:DIII_crossing}. Note that staggering in the $z$-direction destroys all of the invariants of Eqs.~\eqref{eq:q_diii_2z}, \eqref{eq:q_diii_1x}, and \eqref{eq:q_diii_1y}, turning the entire phase diagram into a topologically trivial insulator.

\section{Extended topological classification}
\label{sec:stpt}

In the previous Sections we have presented models showing topological phase transitions protected by average reflection symmetry, which we dub statistical topological phase transitions, following nomenclature of Ref.~\onlinecite{Fulga2014}.
Since the strong index remains constant across these transitions, we need to extend the topological group structure of the periodic TI table in order to properly label the protected phases.
In this Section, we discuss this extension in the context of the models presented above, and show how it applies to systems of any dimensionality and symmetry class.

The phase diagram of the 2d system, Fig.~\ref{fig:pd_D}, has two statistical topological phase transitions. The lower one, $\mu=2$ and $t_d=1/4$, happens at a vanishing Chern number, $\nu=0$. The corresponding phases are a trivial system ($\nu=\nu_y=0$), $\mu>2$, and a WTI ($\nu_y=1$) for $\mu<2$. As such, its robustness to disorder can be understood in the language of Ref.~\onlinecite{Fulga2014}, namely in terms of the different edge localization properties of the two phases.
In the trivial phase the edge is localized: its thermal conductance $G\sim\exp(-L/\xi)$ decays exponentially as a function of system size $L$,  with the localization length $\xi$.
The WTI on the other hand has edge states which avoid localization even in the presence of disorder.
They form so-called Kitaev edges,\cite{Diez2014} characterized by a super-Ohmic conductance $G\sim\sqrt{l/L}$ (with $l$ the mean free path), which scales in a way typical for disordered one-dimensional systems at a critical point.
\cite{Brouwer2000, Brouwer2003, Motrunich2001, Gruzberg2005} Due to bulk-boundary correspondence, the difference in edge localization properties implies that the two phases are topologically distinct, explaining the phase transition's robustness to disorder.

The situation is different for the upper crossing in Fig.~\ref{fig:pd_D}, at $\mu=1$ and $t_d=1/2$. On both sides the strong topological invariant is $\nu=-1$, and as such all edge states avoid localization in both phases. In fact, the thermal conductance of the edge is identical in both systems, $G=|\nu| G_0=G_0$, so the above argument cannot be applied.

\begin{figure}[tb]
 \includegraphics[width=0.9\columnwidth]{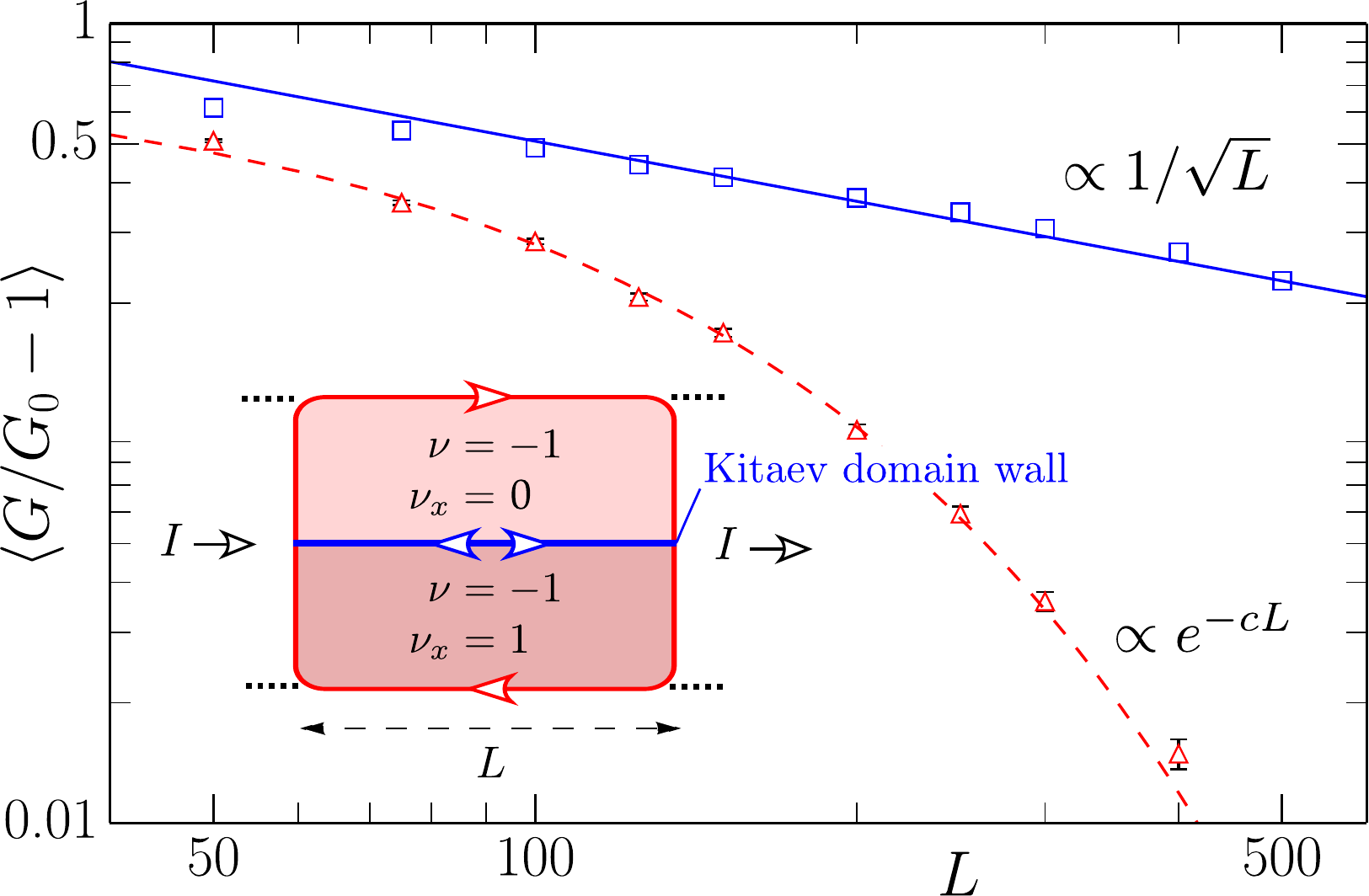}
 \caption{Conductance through a Kitaev domain wall as a function of its length, with and without average reflection symmetry (blue solid and red dashed, respectively). The inset shows the measurement setup, in which conductance flows both through the domain wall and the chiral Majorana edge modes. The quantized edge mode contribution has been subtracted from the plot (vertical axis label). Both the top and bottom halves are described by Eq.~\eqref{eq:HclassD}, using $\mu_{\rm top}=1.5$ and $\mu_{\rm bottom}=0.5$, and keeping all other parameters the same as in Fig.~\ref{fig:pd_D}.\label{fig:interface}}
\end{figure}

Instead, we look at the localization properties of an interface formed between them. Consider a one-dimensional domain wall formed between systems in the two phases ($t_d=1/2$, $\mu<1$ and $\mu>1$). The key observation is that if one of the weak indices differs, the corresponding interface between two strong TIs will behave like the edge of a WTI -- in this case a Kitaev edge, or rather, a \emph{Kitaev domain wall}. Since the index $\nu_x$ changes, the interface parallel to the $x$-direction avoids localization as long as average reflection symmetry is preserved (see Fig.~\ref{fig:interface}). The mobility gap must close along this interface, showing that the two phases are topologically distinct. Therefore, ARS protects weak invariants also when the strong index is nonzero, leading to a $\mathbb{Z} \times {\mathbb{Z}_2}^2$ classification for disordered class D systems in two dimensions.

The situation is similar for the 3d model in class DIII, whose phase diagram is shown in Fig.~\ref{fig:DIII_crossing}. At $t_d=1/4$, the systems goes from a WTI with $Q_1^y=1$ to a trivial insulator as a function of $\mu$, so the different surface localization properties of the two disordered phases imply they are topologically distinct. At $t_d=1/2$ on the other hand, the effect of 1d invariants is obscured by the 2d non-trivial invariant $Q_2^z$, which makes all side surfaces delocalized.
As before, robustness of the topological phases on either side of the crossing can be determined by considering an interface between them.
Our simulations indicate that in this case the interface avoids localization, such that the two phases cannot be continuously connected without closing the mobility gap.

\begin{figure}[tb]
 \includegraphics[width=0.9\columnwidth]{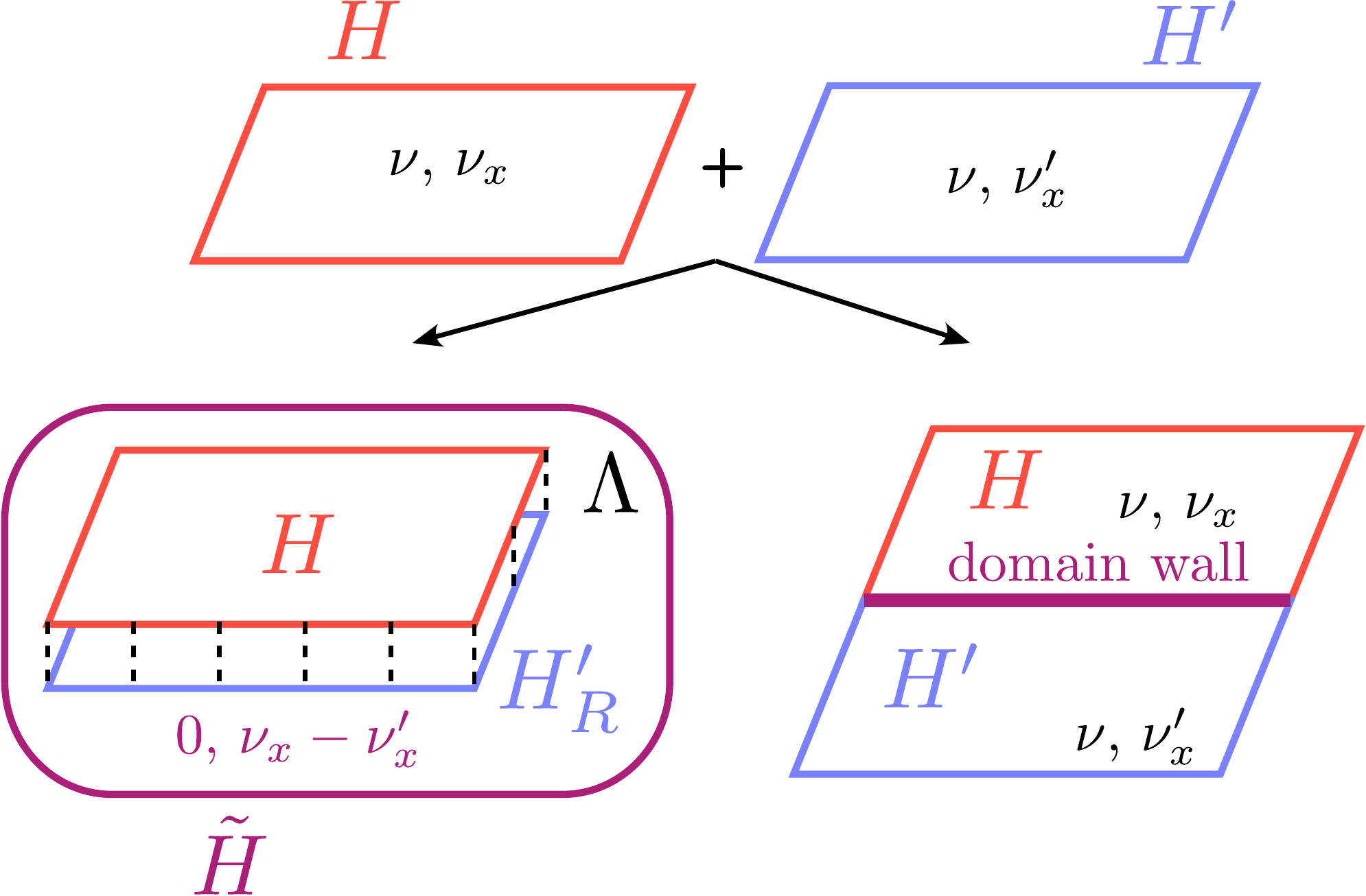}
 \caption{We consider two systems with the same strong indices $\nu$, but different weak indices $\nu_x$ and $\nu_x'$ corresponding to the Hamiltonians $H$ and $H'$. We combine them in one of two ways: on the left we invert the invariants of the second system to $H'_R$ with indices $-\nu$ and $-\nu_x'$ and combine it with the first system using Eq.~\eqref{eq:H_aux}. We make the coupling matrix $\Lambda$ local and having support throughout the bulk of both systems. The combined system has indices $0$ and $\nu_x - \nu_x'$ making it non-trivial only in the weak sense. On the right we put the two systems together with a coupling only over their common edge. Then a weak domain wall is formed with gapless states protected by the non-zero difference $\nu_x - \nu_x'$. This is the generalization of the Kitaev domain wall introduced earlier.
 \label{fig:stacking}}
\end{figure}

In general, strong and multiple generations of weak invariants may affect the localization properties of states at the same boundary. However, contributions of different indices can always be isolated by forming
an interface between two phases with only one index changed.
This is, in fact, analogous to studying the boundaries of a system which is only non-trivial with respect to that particular invariant (see Fig.~\ref{fig:stacking}).

For a $d$-dimensional Hamiltonian $H$, the robustness of one of its topological indices can be determined by studying an auxiliary Hamiltonian in the same symmetry class:\cite{Teo2010, Wen2012}

\begin{equation}\label{eq:H_aux}
 \widetilde{H} = H\oplus H_{\rm R}' \equiv \begin{pmatrix}
                  H & \Lambda \\
                  \Lambda^\dag & H_{\rm R}'
                 \end{pmatrix},
\end{equation}
with $\Lambda$ a symmetry preserving coupling matrix. We choose $H_{\rm R}'$ such that only the nonzero index of interest of $H$ is also nonzero in $\widetilde{H}$ and all other indices of $\widetilde{H}$ trivial. This allows us to use the results of Ref.~\onlinecite{Fulga2014} to show that the boundaries of the auxiliary Hamiltonian \eqref{eq:H_aux} avoid localization in the presence of average symmetries. Therefore, the nonzero index common to both $H$ and $\widetilde{H}$ is robust.

For example, if $H$ is given by Eq.~\eqref{eq:HclassD} with $\nu=-1$, $\nu_x=1$, as happens for $\mu=0$ and $t_d=1/2$, one can choose $H_{\rm R}'$ to have $\nu=1$, $\nu_x=0$, making the combined system\cite{Teo2010, Wen2012, R2010} a WTI only with respect to $\nu_x$. The connection between the Kitaev domain wall formed at the interface between two strong TIs and the auxiliary Hamiltonian introduced in Eq.~\eqref{eq:H_aux} is summarized in Fig.~\ref{fig:stacking}. The combined Hamiltonian can be visualized as the system in the inset of Fig.~\ref{fig:interface}, where the two halves touching at the domain wall have been folded on top of each other. The Majorana edge modes become counter-propagating after folding, such that $\nu=0$, and the domain wall in the original setup becomes the boundary of the folded system. As such, in the following we will restrict ourselves to boundary localization properties, with the understanding that the same results will be reached when multiple non-trivial invariants coexist, either by considering interface properties, or auxiliary Hamiltonians of the form \eqref{eq:H_aux}.

Before proceeding to extend the table of topological insulators to the case where average reflection symmetry is preserved, we shortly review the results of Ref.~\onlinecite{Fulga2014}. We give here only a brief summary, expressed in the language of a concrete physical example, and refer the reader to that paper for the full, detailed derrivation.
This discussion is necessary in order to distinguish between $\mathbb{Z}$ and $\mathbb{Z}_2$ weak invariants.

In the absence of disorder, WTIs have gapless boundary states. They can be thought of as systems formed of weakly coupled layers, where each one caries a strong lower dimensional invariant. Depending on whether the layer index is $\mathbb{Z}$ or $\mathbb{Z}_2$, we consider two constructions: adjacent layers can either have the same value of a $\mathbb{Z}_2$ index, or opposite $\mathbb{Z}$ invariants, $Q$ and $-Q$. A 3d example of the former is a stack of weakly coupled quantum spin Hall systems,\cite{Ringel2012} while the latter is an anti-ferromagnetic stack of quantum Hall systems.\cite{Mong2010, Baireuther2014} In each case, dimerization of the layers can gap out the boundary states, but this is forbidden by exact reflection symmetry.

Note that one can also consider stacked systems in which each layer has the same value of a $\mathbb{Z}$ invariant. In this construction however, the boundary cannot be gapped irrespective of lattice symmetries, so we will not discuss it in the following.

\begin{figure}[tb]
 \includegraphics[width=\columnwidth]{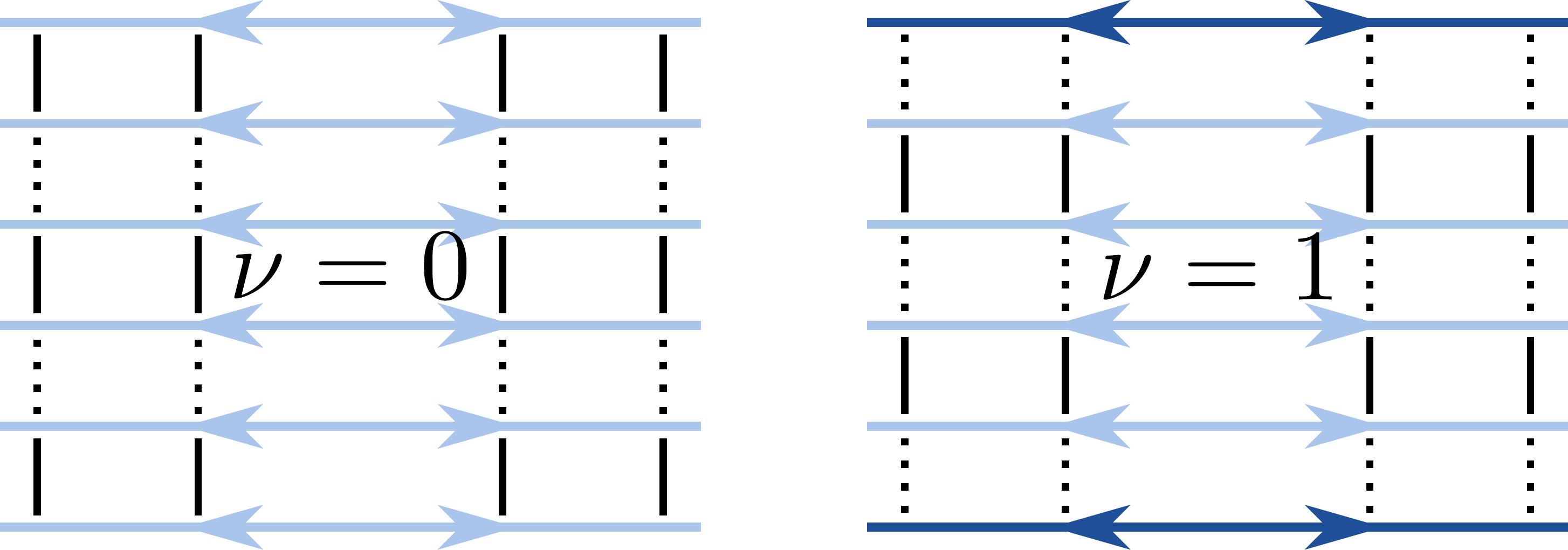}
 \caption{Surface of a stack of quantum spin Hall layers. Horizontal arrows denote the helical edge modes of each layer, and solid/dotted lines indicate strong/weak inter-layer coupling. Reflection symmetry about one layer can be broken in two different ways (left/right panels), leading to different surface invariants $\nu$. On the left the surface is gapped and trivial, whereas on the right the reflected configuration of inter-layer coupling leaves helical edge modes on the surface boundaries (dark color), signaling a non-trivial surface invariant $\nu=1$.\label{fig:refl}}
\end{figure}

When disorder is added, reflection symmetry is explicitly broken, becoming instead an average symmetry of the disordered ensemble. Let us use the stack of coupled quantum spin Hall systems as an example, and assume that the gapless surfaces protected by exact reflection symmetry do indeed become gapped once disorder is introduced. In the presence of a surface gap, we can define surface topological invariants for all elements of the disordered ensemble. Since in 2d (and in general in all dimensions $d\geq 1$) the topological invariant is a self-averaging quantity, it should have the same value for any surface as it does for its reflected counterpart. However, there are two distinct ways of breaking reflection symmetry on the surfaces of a stack of quantum spin Hall layers, with surface invariants that differ by an odd amount, as shown in Fig.~\ref{fig:refl}. Disorder which respects ARS is equally likely to break reflection symmetry in either of the two ways, seemingly contradicting the self-averaging nature of the topological index. The only resolution to this apparent paradox is to invalidate the original assumption, that of a gapped surface.

Ref.~\onlinecite{Fulga2014} showed that boundary states avoid localization whenever the average symmetry changes surface invariants by an odd amount, resulting in a new class of topological phases: statistical topological insulators. With average reflection symmetry, this happens for layered systems in which each layer has a strong $\mathbb{Z}_2$ index, since a change of a $\mathbb{Z}_2$ number can only be odd. Additionally, it was shown this happens for layers with an alternating $\mathbb{Z}$ index $\pm Q$, whenever $Q$ itself is odd. As such, both cases lead to a weak invariant of the disordered bulk system which is $\mathbb{Z}_2$.

The weak invariants found to survive disorder according to the above arguments can then be used iteratively to extend the classification to higher dimensional systems. This is done by studying a system in the same symmetry class but one dimension higher, and considering odd changes in the \emph{weak} surface invariants. Then, the same procedure leads to second generation statistical topological insulators, such as the phase appearing at $\mu=t_d=0$ in the DIII model (Fig.~\ref{fig:DIII_crossing}). The simultaneous presence of two independent average reflection symmetries is required in this case: one guarantees the existence of a weak surface invariant, while the second one changes the value of this weak invariant by an odd amount. Therefore, each strong index, $\mathbb{Z}$ or $\mathbb{Z}_2$, gives rise to infinitely many higher dimensional $\mathbb{Z}_2$ statistical topological insulators in the same symmetry class, which require a larger number of average symmetries for larger dimensionality of the system.

So much for the summary of Ref.~\onlinecite{Fulga2014}. We extend its conclusions to the present case, when multiple invariants coexist. For a $d$-dimensional system in any symmetry class, the classification due to the strong invariant, if any, is extended by each non-trivial invariant of lower dimension, $d' = d-k$, as
\begin{equation}\label{eq:group_extension}
{\mathbb{Z}_2}^\alpha, \quad\quad \alpha={\begin{pmatrix}
                 N \\
                 k
                \end{pmatrix}
},
\end{equation}
where $\alpha$ is a binomial coefficient and $N \leq d$ is the total number of average reflection symmetries. The binomial coefficient in Eq.~\eqref{eq:group_extension} is reminiscent of that found for systems in the absence of disorder,\cite{Kitaev2009, Wen2012} with some important differences. First, it does not go up to the full dimension of the system, but rather to the number of average reflection symmetries which protect the invariants. Second, only $\mathbb{Z}_2$ groups appear, irrespective of whether the lower dimensional index is $\mathbb{Z}$ or $\mathbb{Z}_2$. Lastly, the extension only involves invariants in dimensions $d > d' > 0$, since in zero dimensions the topological invariant is not a self-averaging quantity, making the results of Ref.~\onlinecite{Fulga2014} inapplicable.

We assemble the resulting classification into a new table of topological insulators, which is now no longer periodic, but shows an exponential enlargement of groups with the number of spatial dimensions (see Table~\ref{tab:ti}). In two dimensions we recover the result of Section \ref{sec:model_D} for class D, with a group structure $\mathbb{Z}\times{\mathbb{Z}_2}^2$. In 3d class DIII (Section \ref{sec:model_DIII}), the group is $\mathbb{Z}\times{\mathbb{Z}_2}^6$ with ARS along all directions: there is one integer valued strong index, three 2d weak indices, and three second generation, 1d invariants. If ARS is broken along one direction, by staggering the system for instance, the group becomes $\mathbb{Z}\times{\mathbb{Z}_2}^3$ instead. In that case, only two 2d invariants and one 1d weak index survive.

The extended classification of Table \ref{tab:ti} applies not only to bulk Hamiltonians, but also to Teo and Kane's classification of topological defects,\cite{Teo2010} enabling us to distinguish between \emph{strong} and \emph{statistical} topological defects. An example of the latter is in fact shown in Fig.~\ref{fig:interface}. It's the Kitaev domain wall, a one-dimensional topological defect protected from localization by ARS.

\begin{figure}[tb]
 \includegraphics[width=0.95\columnwidth]{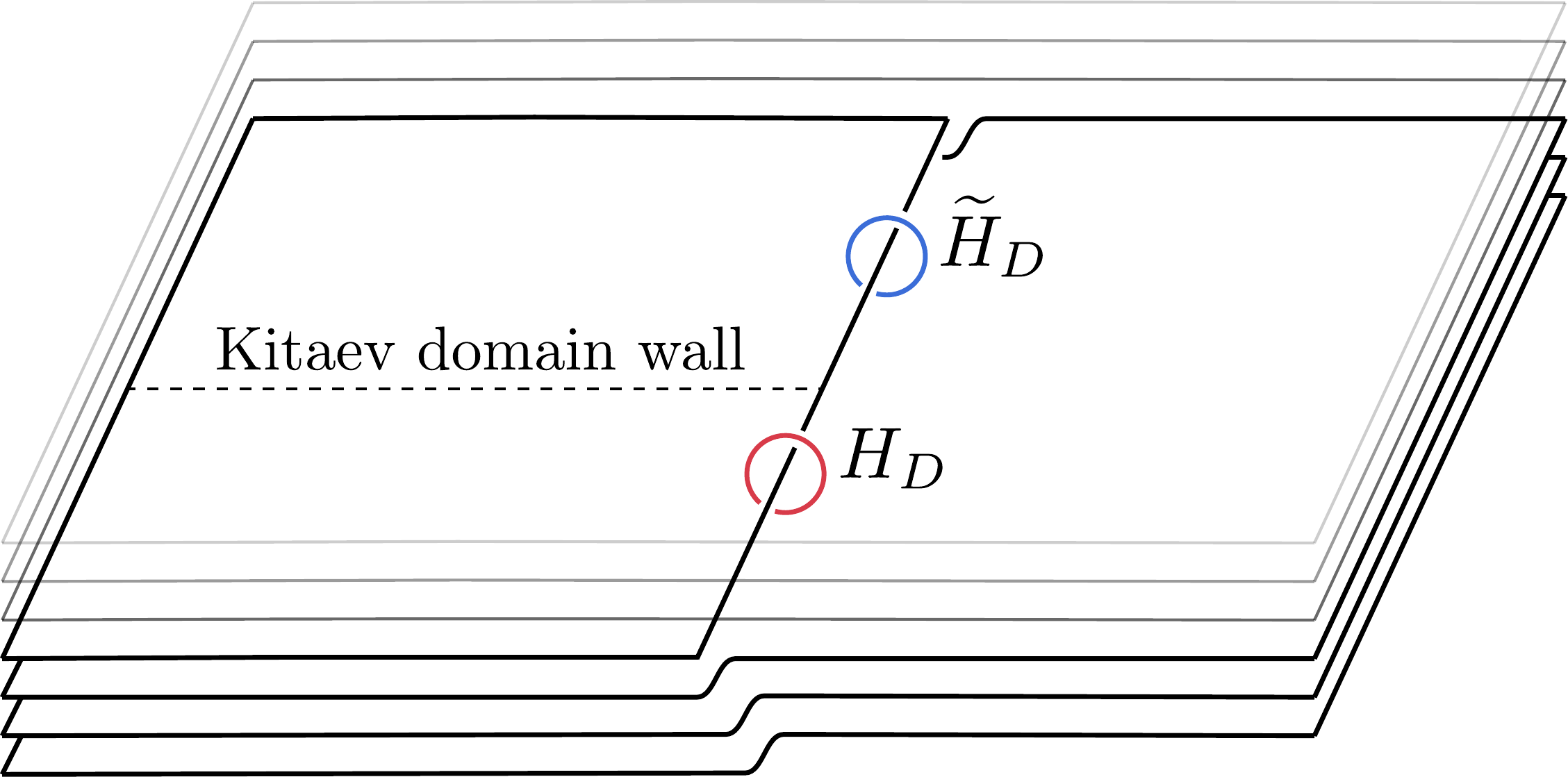}
 \caption{One-dimensional topological defect embedded in a three-dimensional bulk, such as the Hamiltonian \eqref{eq:HDIII} or stacked copies of \eqref{eq:HclassD}. At some point along the defect one of its weak invariants changes, leading to the formation of a Kitaev domain wall. The defect Hamiltonians $H_D$ and $\widetilde{H}_D$ have the same strong invariant, but cannot be deformed into each other without closing a gap, due to the presence of ARS.\label{fig:defects}}
\end{figure}

Since topological defects are classified in terms of the topological properties of Hamiltonians surrounding the defect, they share the same extended group structure as bulk Hamiltonians. Therefore, statistical topological phase transitions in which the strong defect invariant does not change are possible. By using the same interface construction as before, Fig.~\ref{fig:interface}, one can understand these transitions in terms of the properties of the Hamiltonians surrounding them. We show an example in Fig.~\ref{fig:defects}, where the Hamiltonians surrounding two defects with the same strong invariant cannot be adiabatically deformed into each other, since they differ in one of their weak invariants.

\section{Conclusion}
\label{sec:conc}

We have shown how the topological structure of single-particle systems is enhanced by the presence of average symmetries. For concreteness, we have focused on protection due to average reflection symmetry in the presence of disorder, a situation which occurs naturally in many condensed matter systems. We have found that all weak invariants of lower dimensions $d \geq 1$ contribute to the classification at the same time, leading to a group structure which grows exponentially with the number of dimensions.

In general, when multiple invariants affect the localization properties of the same boundaries, the effect of average symmetries can be treated with the construction of Eq.~\eqref{eq:H_aux}, or by forming interfaces between systems. This enables the robustness of each invariant to be studied independently of the others.

Since we focus on the effects of disorder, our classification scheme is different from, and applies also to existing works which generalize the periodic TI table. The same arguments can be applied to any symmetry compatible with the criterion of Ref.~\onlinecite{Fulga2014}. In particular, one may consider instead rotational symmetry, which has also been shown to lead to topologically non-trivial phases and defects.\cite{Jadaun2013, Teo2013, Benalcazar2014} Here too the inclusion of disorder would result in an average rotational symmetry, extending the topological group structure in a similar fashion. This opens possibilities for numerous theoretical studies and widens the possibilities for the experimental observation of the suggested effects.

We have also discussed some of the physical consequences of the extended classification. It can lead to statistical topological phase transitions, governed only by a change in one of the weak invariants. In the presence of average symmetries the bulk gap must close at the transition, even if the topological insulators on either side have the same boundary conductance. Additionally, the extended classification can lead to statistical topological defects, which host gapless modes that are robust to disorder, but which could not exist in the absence of average symmetries.

\acknowledgments

The authors thank T.~Hyart, A.~R.~Akhmerov, and C.~W.~J.~Beenakker for useful discussions. We acknowledge support by the Foundation for Fundamental Research on Matter (FOM), the Netherlands Organization for Scientific Research (NWO/OCW), and an ERC Synergy Grant. DIP additionally thanks NSERC, CIfAR and the Max Planck-UBC Centre for Quantum Materials for support. ICF also acknowledges support from the European Research Council under the European Union's Seventh Framework Programme (FP7/2007-2013) / ERC Project MUNATOP, the US-Israel Binational Science Foundation, and the Minerva Foundation.

\bibliography{crossings}

\end{document}